\documentclass[sigconf, screen]{acmart}

\usepackage{microtype}

\setcopyright{cc}
\setcctype{by}
\acmDOI{10.1145/3832783.3834491}
\acmYear{2026}
\copyrightyear{2026}
\acmISBN{979-8-4007-2882-2/2026/10}
\acmConference[ASE '26]{Proceedings of the 41st IEEE/ACM International Conference on Automated Software Engineering}{October 12--16, 2026}{Munich, Germany}
\acmBooktitle{Proceedings of the 41st IEEE/ACM International Conference on Automated Software Engineering (ASE '26), October 12--16, 2026, Munich, Germany}
\acmSubmissionID{ase26ind-p113-p}
\received{2026-04-30}
\received[accepted]{2026-07-01}

\ccsdesc[500]{Software and its engineering~Acceptance testing}
\ccsdesc[500]{Software and its engineering~Requirements analysis}

\usepackage{amsmath,amsfonts}
\usepackage{algorithmic}
\usepackage{algorithm}
\usepackage[caption=false,font=normalsize,labelfont=sf,textfont=sf]{subfig}
\usepackage{textcomp}
\usepackage{stfloats}
\usepackage{url}
\usepackage{verbatim}
\usepackage{graphicx}

\usepackage{xcolor}
\usepackage{tcolorbox}
\usepackage{booktabs}
\usepackage{listings}
\usepackage{fontawesome5}
\usepackage{hyperref}
\usepackage{enumitem}
\usepackage{balance}

\newcommand{\sectopic}[1]{\vspace{0.2em}\par\noindent{\textit{\bfseries #1}}}

\newcommand{\crv}[1]{\textcolor{black}{#1}}

\begin{document}

\newtcolorbox{rqtakebox}[2][]{
  colback=white,
  colframe=black,
  coltitle=white,
  title=\textbf{#2},
  fonttitle=\bfseries\small,
  fontupper=\small,
  colbacktitle=black,
  left=2pt,
  right=2pt,
  top=2pt,
  bottom=2pt,
  arc=1.5mm,
  boxrule=0.8pt,
  #1
}

\lstdefinestyle{prompt}{
    basicstyle=\ttfamily\footnotesize,
    backgroundcolor=\color{gray!10},
    frame=single,
    breaklines=true,
    keepspaces=true,
    columns=fullflexible,
    captionpos=b
}

\title{Requirements-Augmented Generation for Trustworthy Acceptance Testing of LLM-Based Software}



\settopmatter{authorsperrow=4}

\author{Fanyu Wang}
\orcid{0000-0002-9937-8534}
\affiliation{%
  \institution{Monash University}
  \city{Melbourne}
  \country{Australia}
}
\email{fanyu.wang@monash.edu}

\author{Chetan Arora}
\orcid{0000-0003-1466-7386}
\affiliation{%
  \institution{Monash University}
  \city{Melbourne}
  \country{Australia}
}
\email{chetan.arora@monash.edu}

\author{Zhenping Xie}
\orcid{0000-0002-9481-9599}
\affiliation{%
  \institution{Jiangnan University}
  \city{Wuxi}
  \country{China}
}
\email{xiezp@jiangnan.edu.cn}

\author{Yonghui Liu}
\orcid{0000-0001-7548-5100}
\affiliation{%
  \institution{Australian National University}
  \city{Canberra}
  \country{Australia}
}
\email{yonghui.liu@anu.edu.au}

\author{Kla Tantithamthavorn}
\orcid{0000-0002-5516-9984}
\affiliation{%
  \institution{Monash University}
  \city{Melbourne}
  \country{Australia}
}
\email{chakkrit@monash.edu}

\author{Aldeida Aleti}
\orcid{0000-0002-1716-690X}
\affiliation{%
  \institution{Monash University}
  \city{Melbourne}
  \country{Australia}
}
\email{aldeida.aleti@monash.edu}

\author{Siwei Jiang}
\orcid{0009-0007-2905-276X}
\affiliation{%
  \institution{Tianjin University}
  \city{Tianjin}
  \country{China}
}

\affiliation{%
  \institution{Yiwuyishi Intelligent Technology Co., Ltd.}
  \city{Nantong}
  \country{China}
}

\email{jiangsiwei@tju.edu.cn}

\begin{abstract}

LLM-based software (LBS) integrates large language models as core components to deliver flexible and personalised responses. Unlike traditional software with deterministic outputs, LBSs exhibit context-dependent, stochastic behaviour that renders classical acceptance testing and test oracles insufficient: the same query may require fundamentally different responses depending on users' personas and software context. This gap creates an urgent need for automated acceptance testing frameworks that can autonomously interpret users' instructions, while also raising the challenge of how we can reliably interpret users' intentions in a changing environment. In this paper, we present an automated acceptance testing framework for LBS with calibrated verdict reliability through two technical contributions. First, we introduce \textbf{RE}quirements-\textbf{A}ugmented \textbf{G}eneration (REAG), which interprets user intentions by retrieving relevant software requirements, domain knowledge, and personas via adaptive RAG and self-reasoning to generate context-aware test oracles. Second, recognising that oracle generation may retrieve irrelevant constraints, misinterpret intent, or hallucinate requirements, we introduce a confidence-calibrated cascade judgment that quantifies verdict reliability via simulated expert agreement, accepting high-confidence verdicts, escalating ambiguous cases, or abstaining when uncertain, with empirical reliability guarantees backed by conformal risk control. An industrial case study on a production nutrition advisory application demonstrates that REAG achieves a 3.91/5 oracle quality score, successfully achieving qualified or marginal oracle quality in 82\% of cases. The confidence-calibrated cascade achieves 98.8\% accuracy, improves oracle quality from 3.91 to 4.30 by filtering unqualified outputs, and delivers a 31.7\% cost-efficiency improvement over single-judge baselines, validating industrial viability.

\end{abstract}

\keywords{LLM-based software, Acceptance testing, Test Oracle, Requirements Engineering}

\renewcommand{\shortauthors}{Wang et al.}

\maketitle

\vspace{-1em}

\section{Introduction}
\label{sec:introduction}

Large Language Models (LLMs), with broad applications across a wide range of tasks, are increasingly adopted in modern software as core functional components~\cite{li2024survey,wu2024new,qiu2024llm}. Unlike traditional software, which produces deterministic outputs for identical inputs, LLM-based software (LBS) behaves stochastically and is intent-driven, where the same query may require fundamentally different responses depending on the user's persona, health profile, or operational context~\cite{Bucaioni2025AFSA, Chen2024LLMFMA}. This context-sensitivity is by design, precisely what allows LBS to deliver flexible, personalised experiences across diverse user needs. However, this flexibility comes at a cost to software quality assurance. Traditional testing approaches assume a well-defined action space and predictable program behaviour, where a correct output can be specified in advance and verified by comparison~\cite{goericke2020future, de2023characterizing}. For LBS, neither assumption holds, as what constitutes a correct response is determined not only by functional logic but also by the user's intent, software requirements, and user-specific constraints such as health profiles and usage contexts, a combination that varies with each user and context.

These characteristics have two practical consequences for acceptance testing. First, it is impossible to construct test oracles by requirements coverage alone, since a single user intention may correspond to infinitely many valid LLM responses, making no finite enumeration of expected outputs a reliable oracle. Second, even when oracles can be generated, their correctness cannot be assumed, as oracle generation may retrieve irrelevant constraints, misinterpret user intent, or hallucinate requirements not grounded in actual software specifications. Together, these consequences reframe acceptance testing for LBS as a \emph{requirements-grounded intent reconstruction problem}. User-specific constraints, such as health conditions or usage contexts, are part of the software requirements in LBS and must be treated as first-class inputs to oracle generation, not afterthoughts. This motivates two interconnected objectives:

\begin{itemize}[topsep=2pt, itemsep=0pt, parsep=0pt, leftmargin=10pt]
    \item \texttt{obj-i} \textit{Intention Interpretation.} The system must infer a user's likely intent from an input query and profile, and reconstruct the behavioural expectations the system should have satisfied, subject to software constraints (requirements, persona, and domain knowledge), producing executable pass/fail oracle criteria.
    
    \item \texttt{obj-ii} \crv{\textit{Verdict Reliability.}} Given a generated oracle, the verdict mechanism must reliably determine whether actual system behaviour satisfies the criteria.
\end{itemize}

We present a framework that, rather than enumerating expected outputs, generates executable test oracles by reasoning over software constraints and \crv{calibrates the reliability of verdicts} through a statistically calibrated LLM cascade that simultaneously filters low-quality oracles and reduces evaluation cost. To address \texttt{obj-i}, we introduce \textbf{Requirements-Augmented Generation (REAG)}. Existing RAG approaches retrieve general knowledge to produce better answers to questions. This fails for acceptance testing because the goal is not to answer a question but to specify what the system should have done given a particular user's constraints. REAG addresses this by retrieving software artifacts, specifically requirements, user profiles, and software docs, to reconstruct the behavioural expectations the system should have satisfied, and encoding those expectations as executable pass/fail oracle criteria. The individual components we employ, namely ICRALM scoring~\cite{Rametal2023}, adaptive-k selection~\cite{Taguchietal2025}, and self-reasoning trajectories~\cite{Xiaetal2024}, are established techniques in the NLP domain; the contribution lies in reorienting their collective purpose from knowledge augmentation toward behavioural specification, a goal that none of them individually address. To address \texttt{obj-ii}, we introduce \textbf{confidence-calibrated cascade judgment}. LLM-as-a-judge have been applied to model evaluation and data annotation, but not to acceptance testing, where the judge must serve two simultaneous purposes: assessing whether actual system behaviour satisfies oracle criteria, and detecting whether the oracle itself is reliable enough to trust. Our cascade addresses both by using simulated expert agreement to estimate verdict confidence, accepting high-confidence verdicts, escalating uncertain cases to stronger judges, and abstaining when no judge reaches sufficient confidence. The abstention mechanism acts as a quality gate that prevents unreliable oracle verdicts from propagating. Confidence thresholds are calibrated using conformal risk control~\cite{angelopoulos2022conformal}, providing finite-sample reliability guarantees on accepted verdicts, which, to our knowledge, is the first application of this statistical framework to acceptance testing.

\vspace{-1em}

\subsection{Industrial Context and Motivation}
We evaluate our framework on a nutrition scoring and advisory mobile application developed by Yiwuyishi Intelligent Technology (Nantong) Co., Ltd., a software company with $\approx$20 developers specialising in IT consulting and software outsourcing. The application leverages LLM capabilities to provide personalised nutritional guidance by analysing food packaging images and users' health profiles. The system serves diverse user segments identified through user research, including weight management, athletic performance, and medical conditions such as diabetes, hypertension, and kidney disease. Its core functionality involves OCR-based extraction of nutritional information from food packaging images, retrieval of relevant dietary guidelines from domain knowledge bases, and LLM-powered generation of personalised nutritional assessments. Unlike traditional software, this system must dynamically infer implicit user needs, such as blood glucose management for diabetic users or sodium restriction for those with hypertension, even when these concerns are not explicitly stated in the query. This makes it a representative and challenging case for LBS acceptance testing.

\begin{figure}
    \centering
    \vspace{2mm}
    \includegraphics[width=0.91\columnwidth]{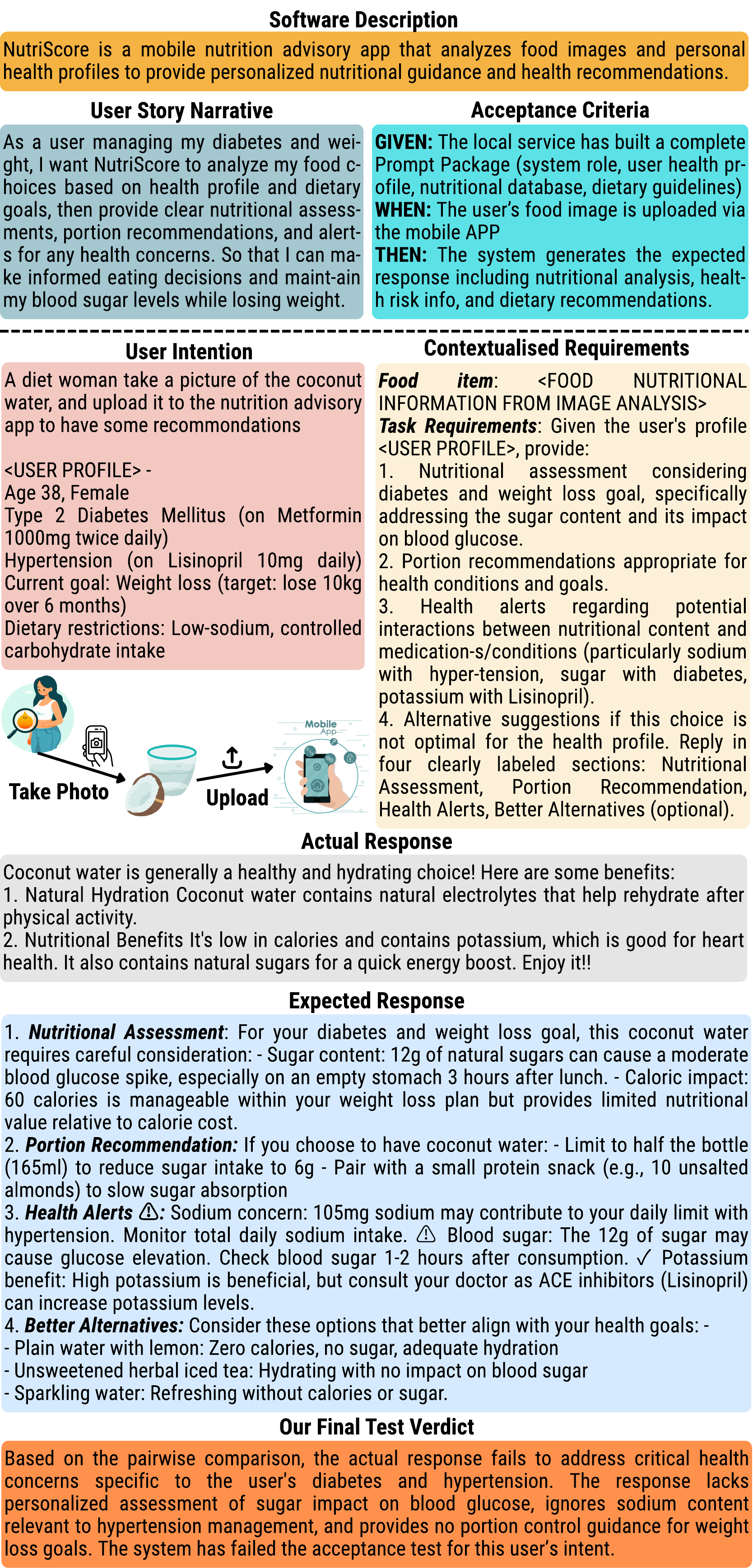}
    \vspace{-0.5em}
    \caption{Acceptance Testing for an LBS (Our case study).}
    \vspace{-1.5em}
    \label{fig:working_example}
\end{figure}

Figure~\ref{fig:working_example} contrasts traditional acceptance testing with our framework on the nutrition app NutriScore. The traditional approach defines Given-When-Then acceptance criteria that verify the system assembles a prompt package and returns a response. While this confirms the service layer behaves correctly, it remains entirely agnostic to whether the LLM's response is appropriate for the user's actual health context. In this example, a diabetic user with hypertension uploads an image of coconut water. The traditional test passes because the system produces a response. However, the actual response provides only generic nutritional benefits and ignores user-specific health concerns. This failure is invisible to traditional testing because the criteria operate at the service layer, not at the semantic layer of the LLM's response.

Our framework addresses this gap by treating oracle generation as the reconstruction of intent from software constraints. REAG retrieves the user's health conditions (Type 2 Diabetes, Hypertension), dietary goals (weight loss), and medication constraints (Metformin, Lisinopril) from the software requirements and persona definitions, then reasons over these constraints to generate a test oracle with explicit pass/fail criteria, such as flagging high sugar content for blood glucose risk and warning about sodium interactions with hypertension medication. The cascade then evaluates the actual response against this oracle, and correctly identifies a test failure that traditional methods would miss entirely. Importantly, if REAG had generated a low-quality oracle, for instance, retrieving irrelevant constraints about athletic performance, the cascade would abstain rather than propagate an unreliable verdict.

\subsection{Contributions}
\begin{itemize}[topsep=2pt, itemsep=0pt, parsep=0pt, leftmargin=10pt]
    \item \textbf{Requirements-Augmented Generation (REAG) for oracle generation}. We reframe test oracle generation as a requirements-grounded intent reconstruction problem. Unlike prior RAG work that retrieves knowledge to answer questions, REAG retrieves software constraints, including requirements, personas, and domain knowledge, to reconstruct what the system should have understood within a user's intent, producing executable pass/fail oracle criteria. \crv{This is the first use} of RAG as a behavioural specification mechanism for acceptance testing, addressing \texttt{obj-i}.
    \item \textbf{Confidence-calibrated cascade judgment as a quality gate}. We introduce a cascade of LLM judges that serves a dual purpose absent from prior LLM-as-a-judge work: validating system behaviour against oracle criteria while screening out unreliable oracles through confidence-based abstention. Thresholds are calibrated using conformal risk control, providing empirical reliability guarantees for accepted verdicts and addressing \texttt{obj-ii}.
    \item \textbf{Industrial evaluation}. We conduct a real case study with industry practitioners on a production nutrition advisory application (346 test scenarios, 46 user profiles), demonstrating 98.8\% accuracy, an oracle-quality improvement from 3.91 to 4.30 through cascade filtering, and a 31.7\% cost reduction relative to a single judge baseline. We note that such industry studies involving real practitioners and proprietary software artifacts in requirements-driven testing are scarce in SE research. \crv{To show the generality, we further discuss in Section~\ref{sec:discussion}.}
\end{itemize}
\section{Background}

\subsection{Retrieval-Augmented Generation}
Retrieval-Augmented Generation (RAG) augments text generation with real-time retrieval from external knowledge sources, reducing hallucinations and outdated information~\cite{Lewisetal2020, Yuetal2024}. However, traditional RAG systems rely on fixed top-k retrieval heuristics that retrieve predetermined numbers of documents regardless of query characteristics, often degrading performance with irrelevant context~\cite{Saxenaetal2025}. Adaptive top-k methods address this by dynamically determining optimal document counts based on query specificity and score distributions, including similarity gap analysis~\cite{Xiaetal2024, Taguchietal2025}, threshold-based methods~\cite{Zhaoetal2024, Caietal2025, Lietal2025}, and ML techniques~\cite{Lietal2024, Saxenaetal2025}.

Recent work has further extended RAG with reasoning-aware retrieval strategies. Self-Reasoning~\cite{Xiaetal2024} enhances retrieved documents with explicit reasoning trajectories, improving reliability and traceability. RATT~\cite{zhang2025ratt} and CoT-RAG~\cite{li2025cot} adopt chain-of-thought prompting within the retrieval loop to improve evidence grounding and response faithfulness. These reasoning-aware approaches share the spirit of our self-reasoning mechanism in REAG, but target general knowledge retrieval rather than software constraint retrieval for behavioural specification. In this work, we use three tuning-free retrieval methods specifically adapted for querying software artifacts and reconstructing user intent. Ram et al.~\cite{Rametal2023} introduce ICRALM, which estimates semantic correlations between queries and documents via LLM log probabilities without fine-tuning. 
Xia et al.~\cite{Xiaetal2024} apply self-reasoning to enhance retrieved documents with reasoning trajectories. 
The key distinction of REAG is that retrieval targets software artefacts, e.g., requirements, personas, and domain knowledge, rather than general knowledge, and the output is an executable behavioural specification rather than generated answer.

\vspace*{-0.5em}
\subsection{Automated Acceptance Testing}
Automated acceptance testing determines whether software satisfies acceptance criteria and meets end-user requirements from the customer's perspective~\cite{Bjarnasonetal2016, Macieletal2019, Antonellietal2019}. Various methodologies have emerged to reduce manual effort, including Behavior-Driven Development (BDD)~\cite{Santosetal2024, Raharjanaetal2020}, Model-Based Testing (MBT)~\cite{Macieletal2019, Straszaketal2015}, and Test-Driven Development~\cite{Breurkesetal2022}. However, automating the generation of acceptance testing artifacts from requirements remains fundamentally limited. A recent survey by Wang et al.~\cite{wang2025requirements} catalogues only 156 primary studies between 1990 and 2024, with very few supporting fully automated acceptance testing. Existing approaches face various challenges. The scalability issue in requirements traceability is identified in Corriveau and Shi~\cite{Corriveauetal2013} when linking hundreds of requirements to thousands of test cases. An abstraction gap exists between coarse-grained requirements and detailed tests~\cite{Antonellietal2019}. Practitioners consistently struggle with test-first practices in real-world contexts~\cite{Breurkesetal2022}. Recent AI-powered approaches~\cite{Ferreiraetal2025, Karpurapuetal2024, Wangetal2023} represent promising directions, with Wang et al.~\cite{wang2025multi} leveraging RAG to generate acceptance criteria from multi-modal requirements. But these works still focus on traditional software rather than LBS.

Beyond these, recent works have proposed comprehensive datas-ets or benchmarks for agents or LLMs, including hallucination~\cite{bang2025hallulens}, safety~\cite{abdullah2025isafetybench}, or even concrete applications~\cite{jimenez2023swe, wang2026leprec}. These studies evaluate LLM capabilities in isolation rather than LLM behavior within deployed, user-centric software systems. Instead, we target this gap by treating oracle generation as requirements-grounded intent reconstruction, enabling acceptance testing where correct behavior is defined by user context and software constraints rather than functional logic alone.

\vspace*{-0.5em}
\subsection{Trustworthy LLMs in Judging}
\label{sec:selective_eval}
LLM-as-a-Judge is an emerging paradigm that extends large language models to serve as evaluators for open-ended and human-preference-aligned tasks~\cite{zheng2023judging,li2024generation}, with applications in model evaluation, data annotation, and reward modeling~\cite{dubois2023alpacafarm,kim2024prometheus}. While LLMs show promise as scalable alternatives to human annotators~\cite{wang2024human,kim2024meganno+}, their stochastic nature introduces systematic biases, inconsistent judgments, and vulnerability to adversarial inputs~\cite{zhou2025self,tian2025overconfidence}. Recent empirical evidence further shows that LLM judges from related model families exhibit significantly correlated errors, agreeing on wrong answers in a substantial fraction of cases~\cite{tian2025overconfidence}. Our cascade mitigates this by selecting judges from distinct model families (Gemini and GPT), as discussed in Section~\ref{sec:evaluation}.

\textit{Conformal risk control}~\cite{angelopoulos2022conformal} offers a principled solution by providing formal guarantees that expected risk remains below a user-specified threshold~\cite{snell2025conformal}. Given a calibration set, the method identifies a confidence threshold that ensures the true disagreement rate stays below a tolerance level $\alpha$ with high probability $(1-\delta)$:
{
\setlength{\abovedisplayskip}{2pt}
\setlength{\belowdisplayskip}{2pt}
\setlength{\abovedisplayshortskip}{0pt}
\setlength{\belowdisplayshortskip}{0pt}
\begin{equation}
P(f_{LM}(x) = y_{human} \mid c_{LM}(x) \geq \hat{\lambda}) \geq 1 - \alpha
\label{eq:conformal_guarantee}
\end{equation}
}
where $f_{LM}$ denotes the model's prediction, $y_{human}$ represents the ground truth label, $c_{LM}$ is a confidence measure, and $\hat{\lambda}$ is the calibrated threshold. The threshold is determined by computing an empirical risk $\hat{R}(\lambda)$ on calibration data, deriving an exact upper confidence bound $\hat{R}^+(\lambda)$ using binomial statistics, and selecting $\hat{\lambda} = \inf\{\lambda : \hat{R}^+(\lambda') \leq \alpha \text{ for all } \lambda' \geq \lambda\}$ through fixed sequence testing. This approach provides exact, finite-sample guarantees for a single calibrated judge. In our cascade setting, we apply conformal risk control independently per judge tier following Jung et al.~\cite{jung2024trust}, providing per-tier empirical reliability bounds on accepted verdicts. Three practical constraints apply to this setting. First, the calibration guarantees require that calibration and deployment data follow the same distribution, a condition that may be partially stressed by our synthetic persona generation~\cite{barber2023conformal}. Second, per-tier bounds do not compose formally at the system level when adaptive routing is applied across stages~\cite{xu2023two}. Third, selective temperature scaling applied post-hoc further relaxes the formal guarantees for judges with insufficient calibration samples. We address these constraints empirically in Section~\ref{sec:evaluation} by reporting coverage-accuracy trade-offs across $\alpha$ settings, and note that recent follow-up work on conformally-calibrated LLM judges~\cite{badshah2026scope} confirms that the marginal guarantees hold well in practice under the i.i.d. assumption.
\section{Methodology}
\label{sec:methodology}
\begin{figure*}
    \centering
    \vspace{2mm}
    \includegraphics[width=0.95\linewidth]{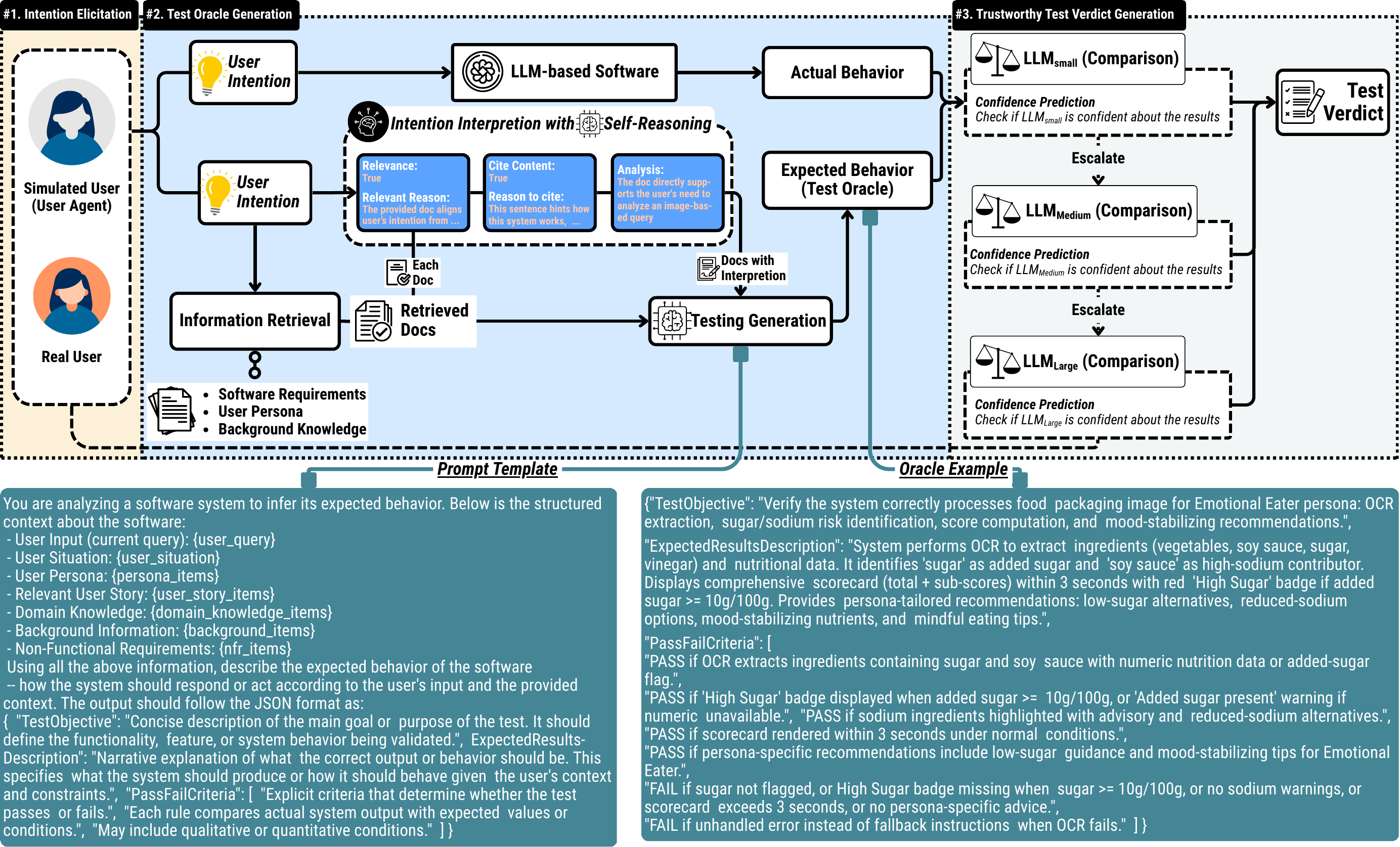}
    \vspace{-0.5em}
    \caption{Overview of Acceptance Testing Framework}
    \vspace{-1em}
    \label{fig:framework}
\end{figure*}

The overview of this framework is exhibited in Figure~\ref{fig:framework}. The framework operates in three stages: (1)~user intention elicitation, (2)~requirements-augmented oracle generation, and (3)~\crv{reliability-calibrated} verdict generation via cascade judgment. For the formulas, we provide a notation summary in Table~\ref{tab:notation}.

\subsection{User Intention Elicitation}
The defining characteristic of LBS is its user-centric behaviour: ``Should I eat this food?'' requires different responses for diabetic, athletic, or pregnant users. This user-dependence is not incidental to the nutrition domain but fundamental to any LBS whose core function is to deliver personalised responses. In a legal advisory system, the same query about contract terms requires different responses for a small business owner versus a multinational corporation. In a fitness coaching system, the same exercise query requires different responses for a post-surgery patient versus a competitive athlete. Consequently, persona $p$ is a first-class input to acceptance testing in any LBS, not a domain-specific design choice: (i) all queries must be associated with personas to determine correct system behavior, and (ii) personas affect query interpretation, e.g., responses for the same query as ``Can I have this \crv{energy} bar?'' would be very different for a diabetic non-athlete versus an athlete.

\begin{table}[t!]
\centering
\caption{Notation Summary}
\vspace{-0.5em}
\label{tab:notation}
\resizebox{\columnwidth}{!}{%
\begin{tabular}{ll}
\toprule
Symbol & Description \\
\midrule
$\mathcal{I}$ & User intention \\
$\mathcal{P}$ & Set of predefined persona categories \\
$\mathcal{Q}$ & Set of real user queries \\
$R, K, p$ & Retrieved requirements, background knowledge, persona description \\
$\mathcal{I}|\langle R, K, p\rangle$ & Interpreted intention with software context \\
$M_i$ & Judge at tier $i$ in the cascade \\
$f_{M_i}(x)$ & Verdict produced by judge $M_i$ \\
$c_{M_i}(x)$ & Confidence score of judge $M_i$ on instance $x$ \\
$\hat{\lambda}_i$ & Calibrated confidence threshold for judge $M_i$ \\
$D_\text{cal}$ & Calibration dataset of oracle-output pairs \\
$\hat{R}(\lambda)$ & Empirical disagreement risk at threshold $\lambda$ \\
$\hat{R}^+(\lambda)$ & Upper confidence bound on empirical risk \\
$\alpha, \delta$ & Risk tolerance and confidence parameters \\
$N, K$ & Number of simulated annotators and in-context examples \\
\bottomrule
\end{tabular}
}
\vspace{-1em}
\end{table}

As LBS operates in a vast, practically non-enumerable space, comprehensive testing requires diverse persona-query combinations. We collect real user queries and define persona categories from actual software requirements. However, within each persona category, users exhibit significant variation, e.g., diabetic users differ in medications, comorbidities, and dietary contexts, all affecting correct system behaviour. Manually collecting profiles covering all such variations is prohibitively expensive. We therefore employ a simulated user agent to systematically expand user profiles within predefined persona categories. The agent $\mathcal{A}_\text{user}$ uses few-shot prompting with real persona categories and queries as examples, generating enriched user situations that capture realistic within-category variation through three mechanisms: grounding in real user queries, constraining to authentic persona categories from software requirements, and learning realistic elaboration patterns from few-shot examples. Specifically, a user intention $\mathcal{I}$ consists of a user query and persona. We generate intentions through:
\begin{enumerate}[topsep=2pt, itemsep=0pt, parsep=0pt, leftmargin=15pt]
    \item Given predefined persona categories $\mathcal{P}$ (from software requirements) and real user queries $\mathcal{Q}$, we randomly sample $\{p_i,\dots,p_j\}$ $\in \mathcal{P}$ and $\{q_i,\dots,q_j\} \in \mathcal{Q}$.
    \item We assign a multi-modal LLM as agent $\mathcal{A}_\text{user}$ with few-shot prompting using the sampled personas and queries as examples.
    \item Agent $\mathcal{A}_\text{user}$ generates enriched user situations within sampled persona categories, producing intentions $\mathcal{I}$ combining real queries with detailed persona profiles.
\end{enumerate}
The generated intentions are combined with real user intentions to ensure comprehensive coverage while maintaining realism.

\subsection{Intent Interpretation with Software Context}
Our framework interprets the user's intent based on the software context by retrieving relevant software requirements. In practice, user queries express intentions implicitly: a user asking ``Help me find a way to the supermarket'' also implicitly requires the nearest location, shortest route, and profile-appropriate transport mode. These implicit requirements must be surfaced for oracle generation. We realise this through three steps: i) ranking relevant software artefacts based on user intention $\mathcal{I}$ using ICRALM~\cite{Rametal2023}, ii) applying an adaptive top-$k$ mechanism to dynamically select relevant documents without manual configuration of $k$~\cite{Taguchietal2025}, and iii) applying self-reasoning to enhance retrieved artefacts with interpretation trajectories~\cite{Xiaetal2024}.

\subsubsection{Relevant Doc Ranking}
ICRALM estimates semantic correlation between a query and document candidates by computing the average log probability on query tokens:
{
\setlength{\abovedisplayskip}{2pt}
\setlength{\belowdisplayskip}{2pt}
\setlength{\abovedisplayshortskip}{0pt}
\setlength{\belowdisplayshortskip}{0pt}
\begin{align}
\text{ICRALM} \!\bigl(\mathcal{I} \mid \text{Task},\text{Doc}\bigr)
&= \frac{1}{|\mathcal{I}|}\sum_{t=1}^{|\mathcal{I}|}
        \nonumber\\
&\quad \log p_{\theta}\!\bigl(i_t \mid \text{Task},\,i_{<t},\,\text{Doc}\bigr)
\end{align}
}
where $\mathcal{I}$ refers to the user's intention, Task is the retrieval prompt, and Doc is the document candidate. Each document is scored and ranked accordingly.

\subsubsection{Adaptive Top-k Selection}
Fixed top-$k$ retrieval risks omitting critical evidence or overwhelming the model with irrelevant context~\cite{Taguchietal2025}. We adopt Adaptive-$k$ retrieval~\cite{Taguchietal2025}, which dynamically determines the optimal number of documents by: (1) computing ICRALM similarity scores, (2) sorting in descending order, (3) computing discrete differences $\text{Gap}_{i}=\text{ICRALM}_{i}-\text{ICRALM}_{i+1}$, and (4) selecting $k = \operatorname*{arg\,max}_{k} \mathrm{Gap}(k)$, where the largest similarity drop marks the boundary between relevant and irrelevant documents. We retrieve the top-$k$ documents plus a small buffer, which is particularly suitable for requirements interpretation where relevant context varies significantly across query complexities.

\subsubsection{Self-Reasoning-based Intention Interpretation}
Traditional R-AG provides retrieved documents without explanation, which may introduce unreliable or untraceable responses~\cite{Xiaetal2024}. We address this through self-reasoning in three steps: i) \textit{relevance checking} that instructs the LLM to judge and explain the relevance of each retrieved document to the user's intention, ii) \textit{evidence checking} that directs the LLM to cite specific sentences from retrieved documents and explain why they support the interpretation, and iii) \textit{application analysis} that synthesizes the previous steps into a concise final interpretation. The retrieved software context consists of requirements $R$, background knowledge $K$, and persona description $p$. We formulate the interpretation as:
{
\setlength{\abovedisplayskip}{2pt}
\setlength{\belowdisplayskip}{2pt}
\setlength{\abovedisplayshortskip}{0pt}
\setlength{\belowdisplayshortskip}{0pt}
\begin{equation}
\mathcal{I}\xrightarrow{{\text{{Reasoning}}}}\mathcal{I}|\langle{{{R},{K},{p}}}\rangle
\label{eq:intention_interpretation}
\end{equation}
}
where $\mathcal{I}|\langle{{{R},{K},{p}}}\rangle$ represents the interpreted intention with the retrieved software context.

\subsection{Test Oracle Generation}
After obtaining the interpreted intention in Eq.~\ref{eq:intention_interpretation}, we generate the expected behaviour of the software as a test oracle. Test cases and acceptance criteria face fundamental limitations for LBS: a single interpreted intention $\mathcal{I}|\langle{{{R},{K},{p}}}\rangle$ may correspond to infinitely many valid LLM responses, making exhaustive enumeration infeasible and exact-matching brittle. Test oracles instead encode executable behavioural properties that validate unboundedly many stochastic responses~\cite{malik2017multi,zhang2025large}, making them the appropriate representation for LBS acceptance testing.

We employ an LLM to generate test oracles from a predefined prompt template, as shown in Figure~\ref{fig:framework}. The oracle specifies a TestObjective, ExpectedResultsDescription, and explicit PassFailCriteria that compare actual system output against expected behaviour. However, test oracles are notoriously difficult to obtain correctly~\cite{Yu2023RetromorphicTA, Jiang2024GeneratingEO, zhang2025large}: LLM hallucinations and interpretation errors from previous steps may produce oracles with incorrect implementations or surface-level descriptions~\cite{farquhar2024detecting,barr2014oracle}. This calls for \crv{an explicit reliability assessment} of the generated oracles.

\subsection{Calibrated Test Verdicts Generation}
\label{sec:verdicts_gen}
The cascade uses three judges: Gemini-2.5-Flash-Lite, GPT-4.1-mini, and Gemini-2.5-Flash, selected from different model families to reduce correlated errors. Recent empirical evidence shows that judges from the same model family tend to inherit shared biases and exhibit correlated errors, agreeing on wrong answers in a substantial fraction of cases~\cite{tian2025overconfidence,kim2025correlated}. Selecting judges from distinct families, specifically Gemini and GPT in our cascade, reduces but does not fully eliminate this risk, as acknowledged in Section~\ref{sec:evaluation}.

\subsubsection{Simulated Annotators for Confidence Estimation}
Given $K$ examples of verdicts from $N$ domain experts, we prompt judge $M_i$ to produce verdict $V$ and explanation $E$, where $f_{M_i}(x)=\langle {V, E} \rangle$, $V \in \{\text{YES}, \text{NO}\}$. We estimate confidence $c_{M_i}(x)$ through simulated annotators by creating $N$ different evaluation perspectives, where each annotator $j$ is conditioned on $K$ in-context examples representing the $j$-th expert's judgment patterns:
{
\setlength{\abovedisplayskip}{2pt}
\setlength{\belowdisplayskip}{2pt}
\setlength{\abovedisplayshortskip}{0pt}
\setlength{\belowdisplayshortskip}{0pt}
\begin{equation}
c_{M_i}(x) = \max_{y} \frac{1}{N}\sum_{j=1}^{N} p_{M_i}(y|x; (x_{1,j}, y_{1,j}), \dots, (x_{K,j}, y_{K,j}))
\end{equation}
}
where $y \in \{\text{YES}, \text{NO}\}$. High confidence ($c_{M_i}(x) \approx 1.0$) indicates simulated experts converge on the verdict, suggesting reliable judgment and well-specified oracle criteria. Low confidence ($c_{M_i}(x) \approx 0.5$) indicates oracle flaws or genuinely difficult conformance judgments. The cascade operates as follows:
\begin{enumerate}[topsep=2pt, itemsep=0pt, parsep=0pt, leftmargin=15pt]
    \item Load the lowest-cost judge as the initial judge $M_0$.
    \item For each PassFailCriterion in the oracle, produce verdict $f_{M_i}(x)$ with confidence $c_{M_i}(x)$.
    \item If $c_{M_i}(x) \geq \hat{\lambda}_i$, accept the verdict and proceed to the next criterion. If $c_{M_i}(x) < \hat{\lambda}_i$, escalate to the next judge $M_{i+1}$.
    \item If no judge produces sufficient confidence, revoke the oracle with status \textit{abstain} for manual expert review.
\end{enumerate}
Abstained cases are collected for human inspection, who can validate oracles or identify defects for iterative pipeline improvement, ensuring only high-confidence verdicts are accepted. In our implementation, we set $N=5$ simulated annotators and $K=3$ in-context examples per annotator. The $K$ examples for each annotator are sampled from the calibration set $D_\text{cal}$ using stratified selection across verdict labels and persona categories, ensuring that each simulated annotator reflects a distinct judgment perspective rather than a homogeneous evaluation pattern. Confidence scores $c_{M_i}(x)$ are obtained by prompting $M_i$ with each annotator's demonstration set and averaging the resulting label probabilities, following the procedure of Jung et al.~\cite{jung2024trust}. When logit-level probabilities are unavailable from black-box APIs, we approximate $p_{M_i}(y|x)$ by sampling $M_i$ with temperature $T=0.7$ across $N$ independent runs and computing the empirical label frequency, consistent with established practice in conformal LLM evaluation~\cite{quach2023conformal}.

\subsubsection{Expert-Calibrated Cascaded Evaluation}
Thresholds $\hat{\lambda}_i$ are calibrated using domain expert annotations based on conformal risk control~\cite{angelopoulos2022conformal}. We construct a calibration dataset $D_{\text{cal}}$ of oracle criteria paired with actual system outputs, annotated by software experts with knowledge of system requirements and domain constraints. Each instance $(x, y_{\text{human}}) \in D_{\text{cal}}$ contains: (i) a test oracle, (ii) actual system output, and (iii) expert verdict $y_{\text{human}} \in \{\text{YES}, \text{NO}\}$. For each judge $M_i$, we calibrate $\hat{\lambda}_i$ using fixed-sequence testing~\cite{bauer1991multiple} as:
\begin{enumerate}[topsep=2pt, itemsep=0pt, parsep=0pt, leftmargin=15pt]
    \item Compute verdict $f_{M_i}(x)$ \& confidence $c_{M_i}(x)$ for each $(x, y_{\text{human}})$ $\in D_{\text{cal}}$.
    \item For candidate thresholds $\lambda$ from high (0.999) to low, compute empirical disagreement risk:
    {
\setlength{\abovedisplayskip}{2pt}
\setlength{\belowdisplayskip}{2pt}
\setlength{\abovedisplayshortskip}{0pt}
\setlength{\belowdisplayshortskip}{0pt}
    \begin{align}
    \hat{R}(\lambda)
    &= \frac{1}{n(\lambda)}
       \sum_{(x,\,y_{\text{human}})\in D_{\text{cal}}} \nonumber\\
    &\quad \mathbb{1}\!\Bigl[
           f_{M_i}(x)\neq y_{\text{human}}
           \;\land\;
           c_{M_i}(x)\ge\lambda
       \Bigr]
    \end{align}
    }
    where $n(\lambda) = \sum \mathbb{1}\{c_{M_i}(x) \geq \lambda\}$ counts instances above threshold $\lambda$.
    \item Compute the exact $(1-\delta)$ upper confidence bound:
    {
\setlength{\abovedisplayskip}{2pt}
\setlength{\belowdisplayskip}{2pt}
\setlength{\abovedisplayshortskip}{0pt}
\setlength{\belowdisplayshortskip}{0pt}
    \begin{align}
    \hat{R}^+(\lambda) &=
      \sup \Bigl\{ R :
      P~\!\Bigl(\text{Bin}\bigl(n(\lambda),R\bigr)
               \le \bigl\lceil n(\lambda)\hat{R}(\lambda)\bigr\rceil\Bigr)
      \;\ge\; \delta
      \Bigr\}
    \end{align}
    }
    \item Select the threshold maximising coverage while bounding risk below $\alpha$:
    {
\setlength{\abovedisplayskip}{2pt}
\setlength{\belowdisplayskip}{2pt}
\setlength{\abovedisplayshortskip}{0pt}
\setlength{\belowdisplayshortskip}{0pt}
    \begin{equation}
    \hat{\lambda}_i = \inf\{\lambda : \hat{R}^+(\lambda') \leq \alpha \text{ for all } \lambda' \geq \lambda\}
    \end{equation}
    }
\end{enumerate}
This guarantees that verdicts accepted by the cascade align with expert judgment at the specified confidence level per Eq.~\ref{eq:conformal_guarantee}, with probability at least $1-\delta$.

In practice, calibrated thresholds may result in imbalanced cascade compositions where certain judges receive insufficient samples (e.g., $n(\hat{\lambda}_i) < \tau$). We apply selective temperature scaling: for each such judge $M_i$, we scale confidence scores as $c'_{M_i}(x) = \sigma(\text{logit}(c_{M_i}(x))/T_i)$ where $T_i > 1$, then re-calibrate $\hat{\lambda}_i$ until all judges handle at least $\tau$ samples (typically $\tau \in [5,10]$), preserving original calibration quality for well-performing judges at $T_i = 1.0$.
\vspace{-0.5em}
\section{Evaluation and Case Study}
\label{sec:evaluation}

\subsection{Research Questions}

We conduct this case study to address three research questions mapping to the two test objectives defined in Section~\ref{sec:methodology}:

\begin{itemize}[topsep=2pt, itemsep=0pt, parsep=0pt, leftmargin=15pt]
    \item \textbf{RQ1 (\texttt{obj-i}):} \textit{How effectively does REAG interpret user intent within software constraints to generate test oracles?} This research question evaluates our framework's ability to achieve \texttt{obj-i}, where user intention $\mathcal{I}$ should remain consistent under both expected and actual software contexts. The reasoning must correctly interpret what users want, understand system limitations, and reason about unstated expectations. We define four evaluation perspectives following prior work~\cite{wang2025multi,arora2024generating}: Relevance, Coverage, Correctness, and Coherence.

    \item \textbf{RQ2 (\texttt{obj-ii}):} \textit{How accurately does the cascade judge determine if system behaviour aligns with interpreted user intent?} This research question evaluates \texttt{obj-ii}, assessing whether the verdict mechanism reliably determines if actual system behaviour satisfies the criteria, while filtering unreliable oracles through confidence-based abstention. We aim to i) screen out cases where oracle interpretation may have failed, and ii) provide statistical guarantees on the \crv{reliability} of accepted verdicts.

    \item \textbf{RQ3:} \textit{What is the cost-efficiency of cascaded evaluation for practical deployment?} We quantify the efficiency of our system by comparing it against the single highest-capability LLM baseline.
\end{itemize}
\vspace{-0.5em}
\subsection{Case Study Design and Protocols}
Our evaluation is structured around three research questions, with metrics tailored to assess oracle-generation quality (RQ1), \crv{verdict selective reliability} (RQ2), and cost-efficiency (RQ3). Following best practices in empirical software engineering research~\cite{runeson2012case,verner2009guidelines}, we employ separate calibration (246 items) and evaluation (100 items) datasets to ensure reliable assessment and prevent overfitting.

\begin{table*}[ht]
\centering
\vspace{2mm}
\caption{Data Categories in Nutrition Advisory Application}
\vspace*{-0.5em}
\label{tab:essential_aspect_4}
\resizebox{\textwidth}{!}{%
\begin{tabular}{p{3.5cm}p{4.5cm}p{3cm}p{3.8cm}p{4.2cm}}
\hline
\textbf{Data Category} & \textbf{Collected Data} & \textbf{Collection Method} & \textbf{Analysis Type} & \textbf{Sample} \\ \hline
D1. Software Artifacts & Requirements docs, domain docs, and persona definitions & Document Analysis & NA & 300+ artifacts \\
D2. Software Behaviours & OCR outputs, parsed data, scores, recommendations, logs & System Execution & NA & 346 scenarios \\
D3. Oracle Evaluation & Multi-perspective experts' evaluation & Interviews \& Forms & Quantitative (Likert Scale) & 100 × 4 scores \\
D4. Verdict Annotation & Test verdicts annotation from experts & Interviews \& Forms & Quantitative (Agreement) & 246 verdicts for calibration \& 100 for evaluation \\
D5. Practitioner Interview & Qualitative discussion about the generated outputs & Interview & Qualitative (Open Discussion) & 30-min interview transcript \\
D6. Extensive Evaluation & Selective agreement, selective confidence, cost comparison & Automated Evaluation & Quantitative (Metrics) & Calibration Acc. on 246 verdicts \& evaluation on 100 verdicts \\ \hline
\end{tabular}
}
\vspace*{-1em}
\end{table*}

\subsubsection{Wohlin's Essential Aspects}
We conduct an embedded single-case study~\cite{yin2018case} on a production nutrition advisory application, following guidelines by Verner et al.~\cite{verner2009guidelines} and Runeson et al.~\cite{runeson2012case}. Our study design explicitly addresses Wohlin's five essential aspects~\cite{wohlin2021case}. \textbf{(1) Empirical Observation:} We conduct systematic data collection across three phases, yielding six complementary data categories (Table~\ref{tab:essential_aspect_4}): software artifacts (300+ artifacts), software behaviours (346 execution traces), oracle evaluations (400 Likert ratings), verdict annotations (346 expert judgments), practitioner interviews (6 transcripts), and automated metrics. \textbf{(2) Contemporary Events:} The application is actively under beta testing (30--50 testers) with all data collection occurring during the study period \textbf{(3) Real-World Context:} We study an actual mobile application with real users, authentic system constraints (API rate limits, credential policies), and 46 user personas identified through user research. \textbf{(4) Multiple Data Collection Methods:} Our triangulation strategy employs document analysis (D1), system execution (D2), Likert evaluation (D3), expert annotation (D4), interviews (D5), and automated metrics (D6), enabling cross-validation where oracle quality issues (D3) can be traced to retrieval failures (D5) and reflected in low verdict confidence (D6). \textbf{(5) Unclear Phenomenon-Context Boundary:} Framework performance is entangled with application context in ways that cannot be isolated experimentally.

\subsubsection{Case Study Protocols}
\textbf{Annotator Recruit Protocols}: All annotators are recruited from the development team of the target software, including six software developers with 3--8 years of experience handling different system functions, ensuring diverse priorities in evaluation. While recruiting from the development team introduces potential familiarity bias, these annotators possess privileged knowledge of system requirements and domain constraints that external annotators would lack, making their judgments more reliable for evaluating requirement-grounded oracle correctness~\cite{runeson2012case}. Also, these developers were not privy to any technical implementation details of our approach to avoid biasing the annotation process. Nonetheless, we acknowledge this in Section~\ref{sec:threats}. \textbf{Annotation Protocols}: We employ two distinct expert annotation sets to establish ground truth. The calibration set uses 246 randomly selected oracle-output pairs, where three experts independently judge conformance (YES/NO) with explanations. Disagreements are resolved through structured post-discussion sessions. The evaluation set additionally uses Likert-scale evaluation collected through interviews with three annotators. Expert annotations reveal \crv{a Fleiss' $\kappa$ of 0.76 (substantial agreement)} across D3, D4, and D5, with near-unanimous agreement after discussion. \crv{To limit leakage between the two sets, the calibration and evaluation sets were annotated in separate sessions, and all calibration labels were frozen before any evaluation-set item was scored or any threshold $\hat{\lambda}_i$ was fit.}
\textbf{Interview Protocol}: We conduct semi-structured interviews with all recruited annotators to gather qualitative insights on oracle quality, \crv{perceived verdict trust}, and framework usability. Each interview lasts 30--45 minutes covering: (i) perceived oracle quality compared to manually-written oracles, (ii) trust in automated verdicts, and (iii) barriers to adoption. Interviews are transcribed into significant points.

\subsection{Evaluation Procedure and Metrics}

\subsubsection{RQ1: Oracle Quality Assessment Metrics}
To evaluate how effectively REAG interprets user intent for acceptance testing within software constraints (\texttt{obj-i}), we conducted a four-dimensional expert evaluation following prior work~\cite{wang2025multi,arora2024generating}:
\begin{enumerate}[topsep=2pt, itemsep=0pt, parsep=0pt, leftmargin=15pt]
    \item \textbf{Relevance} assesses how well the generated oracle aligns with the target software, user query, and situation.
    \item \textbf{Coverage} assesses whether the oracle covers the necessary test objectives \crv{and the persona-based behavioural variations}.
    \item \textbf{Correctness} evaluates the accuracy and soundness of the generated oracle.
    \item \textbf{Coherence} evaluates whether experts can follow the logic of the oracle without additional explanation.
\end{enumerate}
Experts rate oracles on a Likert scale questionnaire~\cite{joshi2015likert}\footnote{From the definition of Likert scale, >= 3/5 are acceptable scores.}. We additionally collect qualitative interview data to understand how interpretation capabilities manifest in practice. To better illustrate the overall score distribution, we also report the average of the four dimension scores (e.g., $\text{average}(4, 3, 2,2)=2.75$) in Figure~\ref{fig:likert}. This average is not an evaluation dimension but a visual reference used in RQ2 to track how cascade filtering affects retained sample quality across $\alpha$ settings.

\subsubsection{RQ2: Selective Reliability Metrics}
By identifying unqualified oracles from RQ1, we evaluate how accurately the cascade judge eliminates them and determines system behaviour alignment with user intent (\texttt{obj-ii}). \emph{Selective Agreement} quantifies the fraction of accepted verdicts that align with expert consensus. High selective agreement (above 95\%) \crv{indicates high conditional reliability on the accepted set; it speaks to the quality of retained verdicts and not to behaviour on abstained or out-of-distribution cases.} \emph{Selection Coverage} quantifies the fraction of oracle criteria for which the cascade produces a definitive verdict rather than abstaining. Higher coverage (above 85\%) means fewer cases require manual review. \emph{Evaluator Composition} analyses the distribution of verdicts produced by each judge tier. We evaluate all metrics across confidence thresholds, plot coverage-accuracy trade-off curves to identify optimal operating points, and compare against single-judge baselines.

\subsubsection{RQ3: Cost-Efficiency Metrics}
To evaluate practical deployment viability, we quantify cost-efficiency using Cost Per Performance (CPP)~\cite{cohn2024costperpoint}, which measures economic efficiency by dividing total computational cost by the percentage of selective agreement.

\subsection{Deployment Details}
We deploy our framework (implemented over three stages) across a mixture of local and API-based environments, with the local environment running on multiple NVIDIA A100-SMI-80GB GPUs. 
\begin{enumerate}[topsep=2pt, itemsep=0pt, parsep=0pt, leftmargin=15pt]
    \item \textbf{\#1 User Intention Elicitation.} To process multi-modal user queries (food images in this application), we employ GPT-5-nano with predefined user personas and nutrition lists in image format. Temperature is set to 0.9 to ensure diversity in the outputs, reflecting the broad variation in real user query styles across persona categories.

    \item \textbf{\#2 Requirements-Augmented Generation (REAG).} We configure the generation in three sub-steps: i) software context retrieval with ICRALM~\cite{Rametal2023}, using GPT-oss-20B~\cite{agarwal2025gpt} as the retrieval kernel (replaceable with lighter alternatives such as SentenceBERT~\cite{reimers2019sentence}); ii) self-reasoning using GPT-5-nano; iii) oracle generation using GPT-5-mini to ensure oracle quality.

    \item \textbf{\#3 Test Verdicts Generation.} Three LLMs form the cascade: Gemini-2.5-Flash-Lite as the lowest judge, GPT-4.1-mini as the mid-tier judge, and Gemini-2.5-Flash as the highest judge. Judge selection prioritises diversity across model families to reduce correlated errors~\cite{tian2025overconfidence,kim2025correlated}, with tiers ordered by increasing capability and API cost as detailed in RQ3.

\end{enumerate}

\subsection{RQ1 Results}

\begin{figure*}
    \centering
    \vspace{2mm}
    \includegraphics[width=0.98\linewidth]{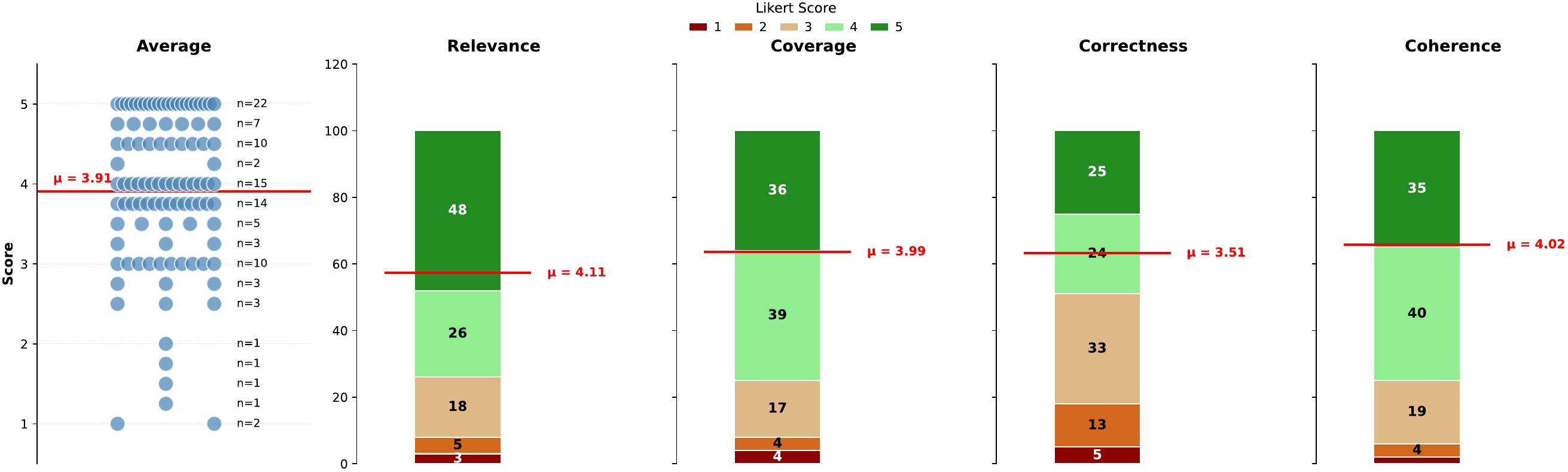}
    \vspace*{-1em}
    \caption{RQ1 - Frequency of Likert Scale ratings for each evaluation criterion by experts.}
    \label{fig:likert}
    \vspace*{-1em}
\end{figure*}

RQ1 examines how effectively REAG interprets user intention in the software context, w.r.t. \texttt{obj-i}. In Figure~\ref{fig:likert}, we present expert evaluation results across four dimensions. The average score for relevance is 4.11/5, followed by coherence (4.02/5), coverage (3.99/5), and correctness (3.51/5), with an overall average of 3.91/5, where the average serves as a composite indicator of oracle reliability rather than a single quality judgment, with dimension-level results reported separately in Figure~\ref{fig:likert}. Experts identified oracles with incorrect implementations and surface-level descriptions, attributing these to hallucinations and incorrect retrieval results.

The distribution across 100 data points reveals that 22 received perfect scores (5/5), while 60 scored between 3.5 and 5.0. \crv{We report two distinct denominators to avoid ambiguity. At the composite level, 7 oracles scored below 3.0 on the four-dimensional average, indicating issues spanning multiple dimensions simultaneously. At the per-dimension level, we regard an oracle as unqualified if it scores below 3 on at least one dimension, which holds for 18 unique oracles; the per-dimension flag counts are 8 for Relevance, 8 for Coverage, 18 for Correctness, and 4 for Coherence, and these sum to more than 18 because a single oracle may be flagged on several dimensions. Oracles scoring exactly 3 on a dimension are marginal (18 for Relevance, 17 for Coverage, 33 for Correctness, 19 for Coherence). The 82\% figure reported throughout is the complement of the 18 unique unqualified oracles.} The defects mainly come from the Correctness dimension.

From interview data, developers highlighted three findings:
\begin{enumerate}[topsep=2pt, itemsep=0pt, parsep=0pt, leftmargin=15pt]
    \item In stage \#1, the simulation agent yields diverse and contextually rich user situations. For example, given ``Coconut Water'' and an aging persona, it generates: ``After a light afternoon walk, the elderly user with T2DM and hypertension sits down for a quick hydration break and chooses coconut water. Contains natural sugars and provides little fiber or protein, so portion size matters to avoid glucose spikes.'' This captures edge cases and enriches sparse queries with contextual intent. \faCommentDots\: \textit{``It came up with situations I didn't even think about, like combining multiple conditions together. That's actually more useful than what we'd write ourselves.''}

    \item In most cases, the self-reasoning mechanism retrieves relevant requirements, but not consistently. Unqualified generations mostly lead to incorrect retrieval, in which the system retrieves irrelevant requirements or omits critical constraints, making the user's intention appear inconsistent between expected and actual contexts. \faCommentDots\: \textit{``When it works, it's great. But sometimes it just grabs the wrong section, like it's looking at a completely different feature, and then everything downstream is wrong.''}

    \item In edge cases, the generation is misled by explicit risks in generated scenarios rather than focusing on core software constraints, causing marginal generations that extend beyond software-defined scope. This occurs when focusing on prominent persona characteristics rather than the requirements defined in the software \crv{artefacts}. \faCommentDots\: \textit{``It picked up on the environmentally friendly angle from the persona and ran with it. That's not something we test for, it's not in the requirements anywhere.''}
\end{enumerate}

The 82\% qualified or marginal rate indicates interpretation succeeds in the majority of cases. The remaining 18\% are attributable to incorrect retrieval and cases where the model follows prominent scenario details outside the defined software requirements.

\subsubsection{Failure Case Analysis} Among the 18 unqualified cases, two distinct failure patterns account for the majority of low scores. \\
(1) REAG retrieves requirements from the wrong functional layer entirely. Several cases involving health-sensitive personas, including a night-shift worker uploading a seasoning ingredient list and a hypertension patient scanning a high-sodium spice product, received oracles whose pass/fail criteria verified service-layer (e.g., OCR) behaviours such as API response handling and OCR confidence thresholds, while omitting the dietary safety constraints central to the user's health context. These scored 1/5 across all four dimensions because they were irrelevant to the user's intent and did not meet any of the required health constraints. This failure (stemming from technical RAG failure) mirrors the traditional acceptance criteria limitation illustrated in Figure~\ref{fig:working_example}, where the system returned a response without assessing its semantic appropriateness.\\ 
(2) reflects scope drift (scores in the 2--3 range) rather than a complete mismatch. REAG retrieves the correct feature area but generates oracles that enumerate too broad a set of system behaviours, diluting the health-critical criteria with generic technical checks. A diabetic user scanning a peanut and wheat sugar product received an oracle that correctly identified sugar threshold rules but also included barcode fallback procedures and carbohydrate normalisation checks, causing Coverage and Correctness to score 2 because the persona-critical constraints were underspecified relative to the surrounding content. A post-surgery recovery patient received an oracle that addressed the right product category but expressed its criteria at the implementation field level rather than in terms of user-observable behaviour. In both patterns, the common thread is that retrieval precision rather than generation quality is the primary bottleneck for oracle correctness.

\begin{rqtakebox}{RQ1 Answer}
REAG achieves an overall oracle quality of 3.91/5 across relevance (4.11), coherence (4.02), coverage (3.99), and correctness (3.51), successfully handling 82\% of cases. The remaining 18\% fail primarily because REAG retrieves technical infrastructure requirements rather than the health and persona constraints that should govern the oracle, confirming that retrieval precision is the primary bottleneck for oracle quality.
\end{rqtakebox}

\subsection{RQ2 Results}

\begin{figure*}
    \centering
    \includegraphics[width=0.97\linewidth]{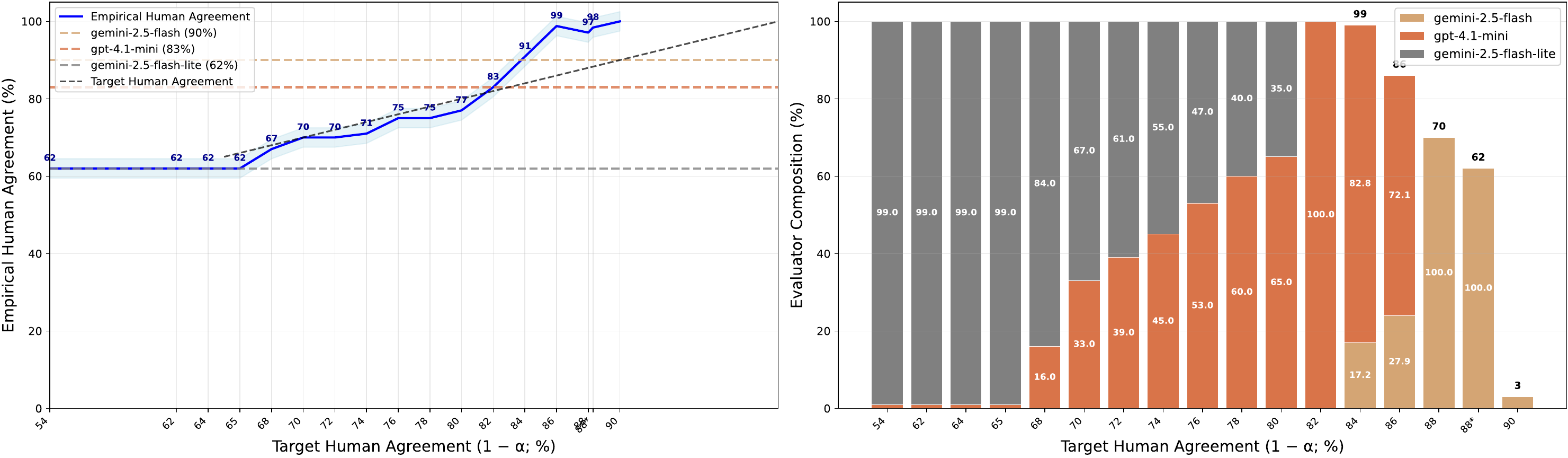}
    \vspace*{-1em}
    \caption{RQ2 - Selection Coverage (R) \& Selective Agreements (L) by $\alpha$ Settings (Default $\delta=0.1$, 88$^*$ is the last observable point).}
    \label{fig:selective_agreements}
\end{figure*}

RQ2 examines whether the cascade judge accurately determines if system behaviour aligns with user intent, relating to \texttt{obj-ii}. The cascade uses three judges: Gemini-2.5-Flash-Lite, GPT-4.1-mini, and Gemini-2.5-Flash, selected from different model families to reduce correlated errors and LLM blind spots~\cite{tian2025overconfidence,kim2025correlated}.

In Figure~\ref{fig:selective_agreements}, the cascade achieves a similar curve to target human agreement when $\alpha \in (0.18, 0.40)$, consistently outperforms when $\alpha \in (0.12, 0.18)$, and overfits when $\alpha \in (0, 0.12)$. At $\alpha=0.14$, the framework achieves 98.8\% accuracy with 14\% abstention. At $\alpha=0.16$, it maintains 90.9\% accuracy with just 1\% abstention. The empirical human agreement curve closely tracks the target across different $\alpha$ values, demonstrating that calibration effectively controls the disagreement rate. Across all $\alpha$ settings, the empirical human agreement meets or exceeds the target threshold $1-\alpha$, confirming that the per-tier conformal calibration controls disagreement risk as intended and validating the finite-sample reliability guarantee in practice. At lower $\alpha$ values, the cascade relies increasingly on higher-tier judges; at higher $\alpha$ values, the lowest-tier judge handles nearly all cases. The sweet spot at $\alpha=0.14$ and $\alpha=0.16$ balances accuracy and coverage across multiple judges.

\begin{figure*}
    \centering
    \vspace{2mm}
\includegraphics[width=0.97\linewidth]{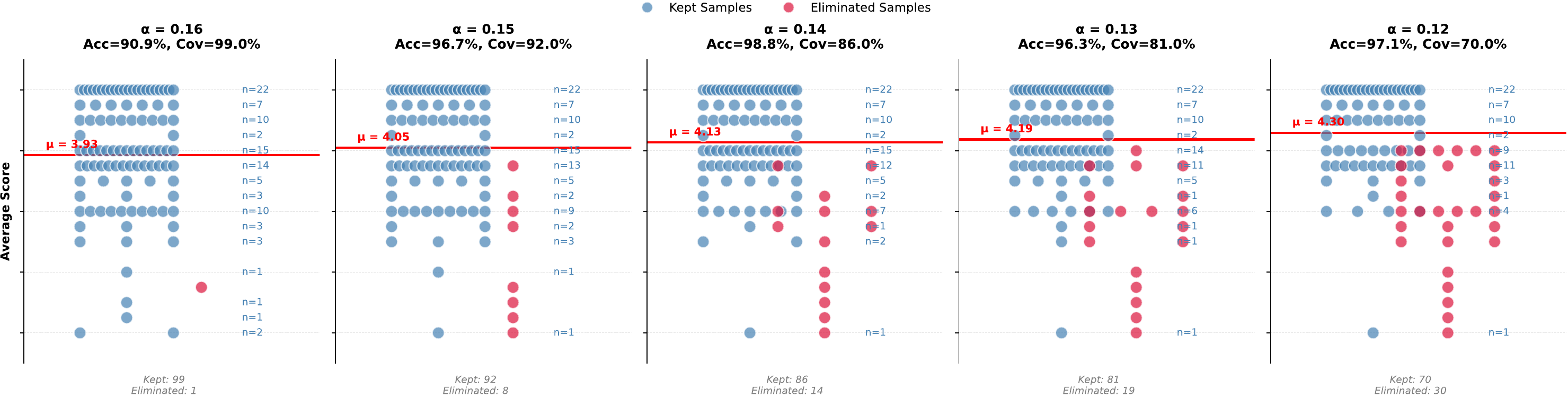}
    \vspace*{-1em}
    \caption{Sample Elimination Trend by $\alpha$ Settings (Default $\delta=0.1$)}
    \label{fig:rq2_elimination}
\end{figure*}

Figure~\ref{fig:rq2_elimination} shows that setting $\alpha=0.13$ eliminates most unqualified generations. The average score of kept samples increases from 3.93 at $\alpha=0.16$ to 4.30 at $\alpha=0.12$, confirming that lower $\alpha$ values filter lower-quality oracles. However, from $\alpha=0.13$ to $0.12$, the method also eliminates qualified results, with coverage dropping from 81\% to 70\%. This over-filtering likely occurs because judges assign high confidence to incorrect verdicts~\cite{tian2025overconfidence}, and properly calibrated thresholds prevent this from affecting the accepted set.

\begin{rqtakebox}{RQ2 Answer}
Our cascade achieves optimal balance at $\alpha=0.14$ with 98.8\% accuracy and 86\% selection coverage, and at $\alpha=0.16$ with 90.9\% accuracy and 99\% selection coverage, outperforming single-model judges (Gemini-2.5-Flash: 90\%, GPT-4.1-mini: 83\%). Kept sample quality improves from 3.93 to 4.30 as $\alpha$ decreases. Thresholds below 0.12 lead to over-elimination due to LLM overconfidence.
\end{rqtakebox}

\subsection{RQ3 Results}

\begin{table}[ht]
\centering
\small
\vspace*{-1em}
\caption{RQ3: Cost per Performance (CPP) Analysis.}
\vspace*{-1em}
\label{tab:cpp_analysis}
\resizebox{\columnwidth}{!}{%
\begin{tabular}{c|ccc|c}
\toprule
\textbf{Setting} & \textbf{Performance} & \textbf{Judge Distribution} & \textbf{Total Cost} & \textbf{CPP} \\
($\alpha$ value) & (Agreement \%) & (J1 / J2 / J3 \%) & (per 100 units) & (Cost/Perf) \\
\midrule
\textit{J2 Baseline} & \textit{83.0\%} & \textit{0 / 100 / 0} & \textit{300.0} & \textit{3.61} \\
\textit{J3 Baseline} & \textit{90.0\%} & \textit{0 / 0 / 100} & \textit{460.0} & \textit{5.11} \\
\midrule
0.36 & 62.0\% & 99 / 1 / 0 & 102.0 & \textbf{1.65} \\
0.18 & 83.0\% & 0 / 100 / 0 & 300.0 & 3.61 \\
0.16 & 90.9\% & 0 / 82.8 / 17.2 & 327.5 & 3.60 \\
\textbf{0.14} & \textbf{98.8\%} & \textbf{0 / 72.1 / 27.9} & \textbf{344.6} & \textbf{3.49} \\
0.12 & 97.1\% & 0 / 0 / 100 & 460.0 & 4.74 \\
\bottomrule
\end{tabular}
}
\vspace*{-1.5em}
\end{table}

In RQ3, we adopt Cost per Performance (CPP)~\cite{cohn2024costperpoint} to quantify cost-efficiency. The three judges correspond to three cost tiers: J1 (Gemini-2.5-Flash-Lite) at base cost (1x), J2 (GPT-4.1-mini) at 3.0x, and J3 (Gemini-2.5-Flash) at 4.6x, based on official API pricing.

Table~\ref{tab:cpp_analysis} shows that the highest-performance setting ($\alpha=0.14$, 98.8\% agreement) achieves a CPP of {3.49}, which is {3.3\% more cost-efficient} than GPT-4.1-mini alone (3.61 CPP) and {31.7\% more cost-efficient} than Gemini-2.5-Flash alone (5.11 CPP), while delivering superior performance. The high-coverage setting ($\alpha=0.16$, 99\% selection coverage) achieves 90.9\% agreement at CPP of 3.60, matching the expensive J3 baseline performance at {29.5\% lower cost}. Settings with high $\alpha$ (e.g., 0.36) are cheapest at 1.65 CPP but produce unacceptable performance (62.0\% agreement) and fail to filter unqualified generations. This confirms the cascade achieves a better cost-performance balance than any single-model approach, making large-scale acceptance testing practical in industrial deployment.

\begin{rqtakebox}{RQ3 Answer}
Our cascade achieves superior cost-efficiency with CPP of 3.49 at $\alpha=0.14$ (98.8\% accuracy), being 3.3\% more efficient than GPT-4.1-mini alone and 31.7\% more efficient than Gemini-2.5-Flash alone, showing that cascaded evaluation reduces cost without sacrificing verdict quality in industrial deployment.
\end{rqtakebox}
\section{Discussion}
\label{sec:discussion}

\crv{Extended from our single embedded case, we further discuss the generality of our framework by separating our claims into three tiers of decreasing generality, and then state concretely what porting the framework to another LBS requires.}

\sectopic{\crv{Tier 1: Mechanisms and settings are domain-independent.}}
\crv{Three findings follow from the architecture of the approach rather than from nutrition. First, persona is a first-class oracle input for any LBS whose core function is personalisation: wherever the correct response is a function of who is asking, an oracle that ignores the asker is underspecified, and the legal-advisory and fitness-coaching cases in Section~\ref{sec:methodology} instantiate the same structure. Second, retrieval precision rather than generation quality is the dominant bottleneck for oracle correctness. This is a property of grounding oracles in retrieved artefacts: a generator cannot specify a constraint it was never shown, so oracle quality is bounded above by retrieval quality in any instantiation of REAG. Our failure analysis supports reading this structurally rather than domain-specifically, since both observed failure modes, retrieval from the wrong functional layer and scope drift, are artefact-organisation failures and should recur wherever technical and user-facing requirements coexist in one corpus. Third, the cascade's cost advantage follows from monotone price--capability tiering among commercial LLMs rather than from any property of our particular judges: so long as cheaper models exist that are adequate for easy cases, routing by confidence dominates always paying for the strongest judge.}

\sectopic{\crv{Tier 2: Configurations are transferable after re-calibration.}}
\crv{The risk tolerance $\alpha$, the per-tier thresholds $\hat{\lambda}_i$, and the assignment of specific models to tiers are \emph{outputs} of the calibration procedure, not constants of the framework. A new deployment re-runs the procedure of Section~\ref{sec:verdicts_gen} against its own expert-labelled calibration set and obtains its own thresholds; our $\alpha=0.14$ operating point carries no authority elsewhere. This is precisely what makes the framework portable while its numbers are not, and it also means the conformal guarantee is re-established locally in each deployment rather than inherited from our study.}

\sectopic{\crv{Tier 3: Measurements that are specific to this case.}}
\crv{The 3.91/5 and 4.30/5 oracle quality scores, the 98.8\% selective agreement, and the 31.7\% cost reduction are joint properties of this application, the quality of its requirements corpus, and this model lineup. They should be read as evidence that the approach is viable in a production setting, not as expected values for other systems. Because REAG's ceiling is set by the artefacts it retrieves over, we expect domains with sparser, less structured, or less consistently maintained requirements to yield lower oracle quality, and domains with mature, well-partitioned specifications to yield higher.}

\sectopic{\crv{What porting the framework requires.}}
\crv{Concretely, an adopter needs four ingredients, none of them nutrition-specific: a machine-readable artefact corpus covering requirements and domain knowledge; a set of persona categories drawn from that corpus rather than invented; a modest expert-labelled calibration set of oracle--output pairs, for which our 246 items are an existence proof rather than a lower bound; and at least two LLM judges at distinct price points, preferably from different vendors. The three pipeline stages themselves are artefact-agnostic. The domains where we would expect the approach to degrade are those lacking the second or third ingredient, that is, where user classes are undocumented or no expert-labelling budget exists, since calibration is what converts an uncalibrated confidence score into a reliability guarantee. Validating these expectations across domains remains future work, and we scope our claim accordingly: we contribute a transferable method together with one industrial demonstration of it, not evidence of cross-domain performance.}

\subsection{Threats to Validity}
\label{sec:threats}
We identify several threats to validity following Wohlin's guidelines~\cite{wohlin2021case}.
\textit{Internal validity} concerns arise from four sources. Annotation inconsistency is mitigated through structured calibration sessions and consensus discussions, achieving \crv{substantial inter-rater agreement (Fleiss' $\kappa=0.76$)}. Annotator bias is partially mitigated through blind annotation protocols, though recruiting from the development team introduces implicit knowledge of system design. Correlated LLM failures are reduced by selecting judges from distinct model families, though cross-family correlated errors remain a known risk~\cite{tian2025overconfidence,kim2025correlated}. Synthetic persona generation may introduce distributional differences between calibration and evaluation data, though empirical results in Section~\ref{sec:evaluation} show coverage guarantees hold across all $\alpha$ settings~\cite{barber2023conformal}.
\textit{External validity} is limited by our single-case study design, which prioritises industrial depth over experimental breadth consistent with embedded case study methodology~\cite{yin2018case}. Generalisation to other LBS domains requires further validation, and domain specialist involvement as co-annotators would strengthen correctness judgments for domain-critical criteria. \crv{Section~\ref{sec:discussion} sets out which claims we believe transfer as mechanisms, which transfer only after re-calibration, and which are specific to this case.}
\textit{Construct validity} threats include the four Likert-scale dimensions that measure expert perception rather than end-to-end fault-detection capability. Connecting oracle verdicts to production defect detection rates would provide stronger evidence and is identified as future work.

\vspace{-0.5em}
\section{Conclusion}
We have presented an automated acceptance testing framework for LLM-based software, addressing the fundamental challenge of testing systems with non-deterministic, context-dependent behaviour through two technical contributions: Requirements-Augmented Generation (REAG) for context-aware oracle generation, and confid-ence-calibrated cascade judgment with conformal risk control for verdicts carrying calibrated selective reliability. The framework achieves two test objectives: interpreting user intent within software constraints (\texttt{obj-i}) and providing statistically reliable verdicts on behaviour alignment (\texttt{obj-ii}). Our industrial case study on a nutrition advisory application with 346 test scenarios demonstrates that REAG achieves an average oracle quality of 3.91/5. The cascade judge achieves 98.8\% accuracy and improves average oracle quality by 10\% through cascade filtering, while being 31.7\% more cost-efficient than the strongest single judge. Per-tier conformal calibration controls disagreement risk as intended, with empirical human agreement meeting or exceeding the target threshold across all $\alpha$ settings.

Our evaluation reveals key limitations: 18\% of oracles are affected by retrieval errors or cases where the model follows prominent scenario details outside the defined software requirements, and edge cases with ambiguous intent still require human judgment. Future work should validate the framework across diverse LBS domains, involve domain specialists as co-annotators, and connect oracle verdicts to production defect detection rates.

\vspace{-0.5em}

\section{Data Availability Statement}
The code and prompt templates are publicly available at~\cite{repo}. The full evaluation dataset cannot be released due to a confidentiality agreement with the industrial partner. To support replication and community use, we provide a set of runnable anonymised data points in the repository.

\balance
\bibliographystyle{ACM-Reference-Format}
\bibliography{references}

\end{document}